\documentclass[nobm,published]{epl}
\usepackage{epsfig}
\usepackage{rotating}
\newcommand{\gl}[1]{eq. (\ref{#1})} 
\newcommand{\gls}[2]{eqs. (\ref{#1},\ref{#2})} 
\newcommand{\glto}[2]{eqs. (\ref{#1}) to (\ref{#2})} 
\def\gtrless{\raise2.5pt\hbox{$>$}\llap{\lower2.5pt\hbox{$<$}}}
\def\gtrapprox{\raise2.5pt\hbox{$>$}\llap{\lower2.5pt\hbox{$\approx$}}}
\newcommand{\bsq}[1]{\begin{subequations}\label{#1}}
\newcommand{\esq}{\end{subequations}}
\newcommand{\beq}[1]{\begin{equation}\label{#1}}
\newcommand{\eeq}{\end{equation}}
\newcommand{\beqa}[1]{\begin{eqnarray}\label{#1}}
\newcommand{\eeqa}{\end{eqnarray}}

\title{Structure and thermodynamics 
of colloid--polymer mixtures: a macromolecular approach}
\shorttitle{Structure of colloid--polymer mixtures}
\author{M.~Fuchs\inst{1} and K.~S.~Schweizer\inst{2}}
\institute{
\inst{1} Physik-Department, Technische Universit{\"a}t M{\"u}nchen,
85747 Garching, Germany\\
\inst{2} Departments of Materials Science \& Engineering and Chemistry,
University of Illinois, Urbana, Illinois 61801
}
\rec{10. April 2000}{}

\begin{document}
 
 
\maketitle

\begin{abstract}
The change of the structure of concentrated colloidal suspensions upon
addition of non--adsorbing polymer is studied within a two--component,
Ornstein--Zernicke based liquid state approach. The polymers'
conformational degrees of freedom are considered
 and excluded volume is enforced at the  segment level. 
The polymer correlation hole,  depletion layer, and 
excess chemical potentials  are described in agreement
with polymer physics theory 
in contrast to models treating the macromolecules as effective  spheres.
Known depletion attraction effects are recovered for low particle density,
while at higher densities novel many--body  effects emerge which become
dominant  for large polymers. 
\end{abstract}

Colloid--polymer mixtures (CPM) play an important role among dispersion systems
for quite different reasons.
On the one hand their technological use has long been realized and the specific
colloid--colloid
interaction caused by free polymer, the depletion attraction
\cite{Asakura,Gast98}, 
is exploited  to induce flocculation or phase separation in dispersions. 
On the other hand, this system allows one to experimentally address the
fundamental question about the requirements on the pair potential 
for  a liquid phase to exist \cite{Gast98,Ilett95}.
 For even more complex systems  CPM serves as a model system in
order to address protein crystallization \cite{zukoski,Poon97} and other
phenomena involving spherical nanoparticles.

Recent field theoretic considerations of a few colloidal particles in dilute
polymer  solutions  \cite{eisenriegler}, and scaling arguments for semidilute
solutions  \cite{joanny}, have clarified the polymer structure close to
particles and provided a fundamental understanding of  the
origins of the depletion attraction. The phase diagram of sterically stabilized
hard--sphere like colloidal particles in polymer solutions close to their
$\Theta$--point has been mapped out in detail \cite{Ilett95}, and a rather
successful mean--field like theory for it exists \cite{Lekkerkerker92}. The
thermodynamic approach of  Lekkerkerker et al. maps the CPM onto a binary hard
sphere mixture with \underline{non}--additive radii, allowing the effective
polymer spheres (EPS) to overlap 
but  excluding them from the
colloids.  Simulation studies for a closely related model
support some of the theoretical predictions \cite{frenkelprl}.  

So far little, however, is known about the structure of the 
CPM, nor how depletion phenomena change when particles are small compared to
polymers.  Whereas neutron scattering
experiments \cite{tong}  on rather low density colloids  could be fit
into an effective pair potential  description, more recent light
scattering experiments on more concentrated systems 
\cite{Moussaid99} have evaded theoretical explanations with  EPS models
\cite{Louis99}, and the assumption of a colloid structure
unperturbed by polymer fails \cite{Moussaid99,Lekkerkerker92}. 
Removing  the assumption of an unaffected local colloid structure forces one to
develop a new approach to CPM which explicitly addresses  local
structure. Moreover, as the packing of polymers into the void space between
 particles is of intrinsic interest on its own, a microscopic two--component
approach appears desirable and shall be presented in this letter.
This macromolecular approach further provides the unique possibility to explore
the full range of colloid--polymer size ratios. 

We consider a two--component system consisting of hard spheres of diameter
$\sigma_c$ at packing fraction $\phi_c=\frac \pi6 \varrho_c \sigma_c^3$, and
 polymers modeled as  chains of  segments (excluded volume diameter
$\sigma_p$) 
characterized by a (Pad{\'e}--approximated)
Gaussian intramolecular structure factor $\omega(q)$ \cite{kcur3}, where
$q$ is the wave vector. In obvious two--dimensional
matrix notation, with a diagonal matrix of intramolecular
structure factors ($\hat{\omega}_{11}\equiv\omega(q)$,  and
$\hat{\omega}_{22}=1$),  diagonal matrix of site number densities,
$\varrho_{ij}$, and a matrix of direct,
$\hat{c}_{ij}(q)$, and intermolecular, $\hat{h}_{ij}(q)$, site--site
correlation functions, the Ornstein--Zernicke--like equations for the total
structure 
factors, $\hat{S}_{ij}(q)$,  are \cite{kcur3,chandl}:
\beq{e1}
\hat{S}^{-1}(q) = [ \varrho \hat{\omega}(q) + \varrho \hat{h}(q) \varrho ]^{-1}
=  \hat{\omega}^{-1} \varrho^{-1} - \hat{c}(q)\; .
\eeq
Together with the constraints of (additive) local steric exclusion,
$g_{ij}(r<\frac 12 (\sigma_i+\sigma_j))=0$, where the  $g_{ij}(r)=1+h_{ij}(r)$ 
are the intermolecular pair correlation
functions, \gl{e1} may be viewed as a definition of the 
effective interactions $c_{ij}(r)$. For the pure colloid component we adopt the
well established Percus--Yevick (PY) approximation, $c_{cc}(r>\sigma_c)=0$.
For the pure athermal
 polymer component, the polymer reference interaction site model
(PRISM) approach is employed as it has proved versatile and
successful for polymers \cite{kcur3,chandl}. PRISM considers
the segment averaged polymer structure and, in the most simple version,
adopts a PY--like closure, $c_{pp}(r>\sigma_p)=0$ \cite{kcur3,chandl}, also
derivable from a Gaussian field theoretic perspective \cite{Chandler93}.

\begin{figure}
\twofigures[scale=0.38]{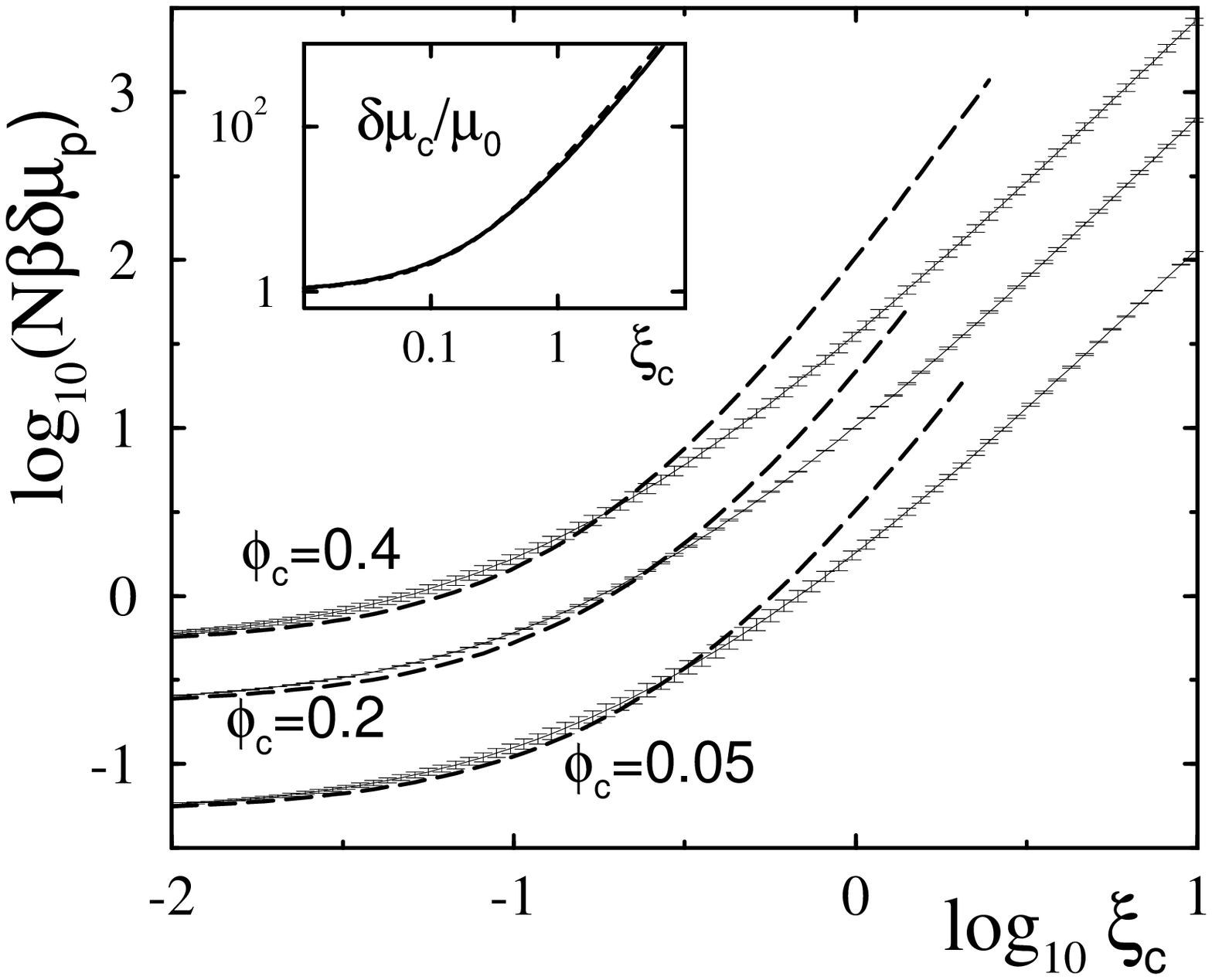}{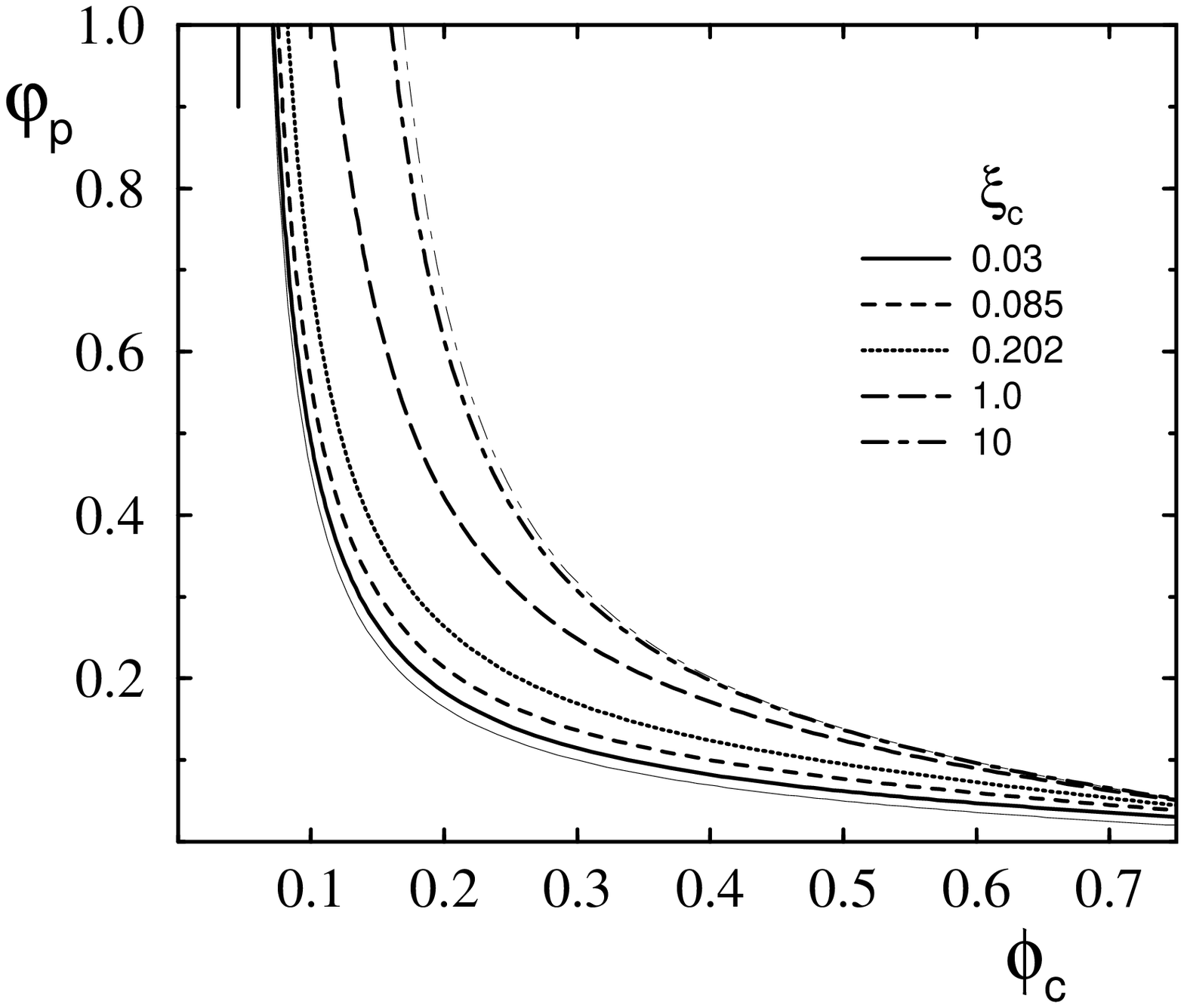}
\caption{Excess chemical potential $N\beta\delta\mu^{\rm (c)}_p|_{\varrho_p=0}$
for inserting one polymer into a hard sphere fluid versus polymer size for
three 
colloid packing fractions. The present results, where  error bars mark
the differences from \protect\gls{e3}{e4},  are compared to
the equivalent PY results for inserting a sphere of radius equal to
$R_g$ (dashed lines) as used in the phantom sphere approach of Ref.
\protect\cite{Lekkerkerker92}. The inset compares
the excess free energy cost (solid line from Eq. (\protect\ref{e5}) where
$\mu_0=\frac\pi6 c_p \sigma_c^3/\beta$) for adding one colloidal sphere to a
dilute polymer solution as a function of relative polymer size,
$\xi_c=R_g/(\sqrt 2\sigma_c)$, with the field theoretic result,  dashed
line  \protect\cite{eisenriegler}.
\label{fig01}}
\caption{Spinodal curves for various size
ratios. The polymer size independent asymptote $\phi_c\to1/22$ for
$\varphi_p\to\infty$ is marked by a vertical bar. The limits for small or large
polymer sizes are included as thin solid or dot--dashed line, respectively. 
\label{fig02}}
\end{figure}

To close the integral equations, a further approximation for the
colloid--polymer direct correlation function is required. We propose a novel
generalization of the PY closure motivated by the known physical behavior of
polymer packing near a spherical particle \cite{eisenriegler}:
\beq{e2}
\hat{c}_{cp}(q)= \frac{\hat{c}^s_{cp}(q)}{1+q^2\lambda^2}\;,
\quad\mbox{with}\quad  c^s_{cp}(r>\frac{\sigma_c+\sigma_p}{2})=0\;.
\eeq
On the segmental level, again the central idea of a short--ranged steric
interaction is used. But, an effective entropic repulsion between segments and
particle is present over a length scale $\lambda$ beyond contact due to chain
connectivity constraints. On physical grounds $\lambda$
cannot be larger than the smaller of the two sizes (colloid or polymer).
In general, the nonlocality length $\lambda$ will
depend on the densities and relative sizes, and additional information is
required  to determine it. We use the argument of thermodynamic consistency
and calculate the excess free energy of insertion per segment of a polymer 
coil into the pure colloidal hard sphere solution via two
independent routes \cite{chandl}. From the compressibility route of
concentrating  the colloid solution around the polymer 
($\beta=1/k_BT$):
\beq{e3}
\beta \delta\mu^{\rm (c)}_p|_{\varrho_p=0} = 
- \int_0^{\varrho_c} d\varrho'_c\;
\hat{c}_{cp}(q=0,\varrho_c')\; ,
\eeq
and from the charging route which considers the free energy
required to grow the  particles 
from  points to their actual sizes:
\begin{eqnarray}\label{e4}
\beta \delta \mu^{\rm (g)}_p|_{\varrho_p=0} = 
\frac{\pi \varrho_c\sigma_c}{2} 
\int_0^1\!\!\!\! d\zeta (\sigma_p\!\!+\!\!\zeta\sigma_c)^2 g^{\rm
(\zeta)}_{cp}(\frac{\sigma_p\!\!+\!\!\zeta\sigma_c}{2}) + 
2 \pi \varrho_c^2\sigma_c 
\int_0^1\!\!\!\! d\zeta (\sigma_p\!\!+\!\!\zeta\sigma_c)^2 
\frac{\partial g^{\rm
(\zeta)}_{cc}(\zeta\sigma_c)}{\partial \varrho_p}|_{\varrho_p=0}\; . 
\end{eqnarray}
Equating the excess free energies from both routes determines $\lambda$, and,
as we first focus 
on dilute polymer solutions,  determining it at $\varrho_p=0$
appears justified. As a technical approximation we analytically determine the
leading limiting behaviors and match the following Pad{\' e}--form to
$\lambda$: $\lambda^{-1}= \xi^{-1} +
\frac{1+2\phi_c}{1-\phi_c}\frac{1}{\lambda_1\sigma_c}$, where
$\lambda_1=(\sqrt{5}-1)/4$.  
Here $\xi$ is the (collective) polymer correlation length and determines the
width of the depletion layer in the large particle limit \cite{joanny,kcur3}. 
For accessible colloid densities and size ratios, the different routes,
\gls{e3}{e4},  to the insertion
free energies at vanishing polymer concentration then agree with  relative
errors smaller than 15 \%.  
Representative results are shown in fig. \ref{fig01}.
Using the results for $\xi$  from \glto{e1}{e4} at finite polymer
concentrations  provides a reasonable approximation to the exact $\lambda$ at
all  densities. The chemical potential  to add single colloidal spheres to a
polymer solution from the compressibility
and free energy routes then agrees within an error of a factor 3 at most.

The aspect of a large length scale separation, $\sigma_c/\sigma_p\approx 10^3$,
thwarts numerical studies of many realistic systems, but can be exploited in a
scaling limit of shrinking the polymer segment size, $\sigma_p$,  to zero; for
a rigorous derivation of this field theory inspired ``thread--limit'' of PRISM
for Gaussian chains see   
\cite{threada}. Then the system is characterized by the 
colloid packing fraction, $\phi_c$, the reduced polymer concentration,
$\varphi_p= 2\pi c_p \xi_c^3$, where $c_p$ is the polymer--molecule number
density,  and the polymer size $\xi_c=R_g/\sqrt2$, where $R_g$ is the radius of
gyration.  Note, that we now use the colloid diameter as unit of length;
 $\sigma_c=1$.   Flexible Gaussian polymers 
are obtained where the statistical segment size $\sqrt{6}\sigma_p$ 
follows from \gls{e3}{e4} for $\varrho_p\to0$ and $\varrho_c\to0$. 
The effective polymer concentration, $\varphi_p\propto c_p/c_p^*$, is measured
relative to the dilute--semidilute crossover concentration, 
$c_p^*$, where the polymer coils start to overlap.
This first result from the present macromolecular approach
 justifies EPS models, where $\eta_p=\frac{4\pi}{3}
c_p R_g^3\approx \varphi_p/0.53$ is used from the outset.
For $\phi_c=0$, the polymer correlation length is
$\xi=\xi_c/(1+2\varphi_p)$ \cite{kcur3}.  
The connection of the thread limit of PRISM  to Gaussian
field theory and scaling approaches has been discussed \cite{kcur3,Chandler93}.

The non--linear integral equations can now be solved with the Wiener--Hopf 
factorization technique; details will be presented elsewhere.
The lines of
spinodal instabilities, where the   (partial) compressibilities diverge
indicating fluid--fluid phase separation, can be found accurately without the
need of extrapolations of numerical integration schemes. In fig. \ref{fig02},
one observes the trend that the required dimensionless polymer density
increases as $R_g$ grows and/or particle diameter decreases.
This prediction \cite{footnote} is  in qualitative agreement with many 
experiments on polymer-- colloid, protein or micelle suspensions
\cite{Ilett95,spinodal}, but in disagreement with EPS models which do not
account for polymer--polymer repulsive interactions, many
body depletion effects, nor particle penetration of polymer coils, all of which
tend to reduce the tendency for phase separation.

The polymer segment 
profile close to a colloidal particle is correctly predicted to be
of a parabolic form:
$g_{cp}(r) \dot{=} \frac{A}{\lambda\xi} (r-\sigma_c/2)^2 + \ldots$, where
$A\to \frac 12 + (\lambda + \xi)/\sigma_c$ for $\phi_c\to0$.
For nonzero $\lambda$, and simplifying to
small polymers, $\xi_c\ll\sigma_c$, the number of polymer segments in contact
with the colloid  particle is of order unity  for a single polymer and becomes
independent of the polymer degree of polymerization in the semidilute regime,
$\varrho_p g_{cp}(r\approx\frac{\sigma_c+\sigma_p}{2})
\propto c_p (1+2\varphi_p)^2 \to (\varrho_p\sigma_p^3)^3$.  This behavior is
in agreement with scaling \cite{joanny} considerations. Quantitatively we find
good agreement as $\varphi_p\to0$:
$A\to1/2$  or $A\to\frac{\xi_c}{\sigma_c}$ for $\xi_c\to 0$ or
$\xi_c\to\infty$,  respectively,
compared to $A\to1/2$  or $A\to4\lambda_1\frac{\xi_c}{\sigma_c}$
from field theory for a single Gaussian chain  \cite{eisenriegler}.
For the PY closure, where $\lambda=0$, this behavior is
violated and thus the polymer induced depletion attraction is significantly 
underestimated \cite{avik98}.  The free energy of insertion of a 
sphere into a polymer solution can be obtained 
from the free energy route (analog of  \gl{e4}) as:
 \begin{eqnarray}\label{e5}
\beta \delta \mu^{\rm (g)}_c|_{\varrho_c=0} = \frac{\pi c_p \sigma_c^3
\xi^2_c}{6\xi^2} 
(1 + \frac{9+36 \lambda_1}{6\lambda_1} (\frac{\xi}{\sigma_c}) + 
\frac{6}{\lambda_1}(\frac{\xi}{\sigma_c})^2 ) \; .
\end{eqnarray}
For dilute polymer solutions,
 \gl{e5}, is compared with field theoretic calculations 
\cite{eisenriegler} in the inset of fig. \ref{fig01}, and excellent  agreement
is found.
Whereas the colloid excess chemical potential
 measures the number of excluded polymer molecules
  for small coils, 
for large polymers, $\xi\gg\sigma_c$,  only segments along a strand
of length proportional to the colloid size need to be rearranged, and the 
result becomes intrinsic (independent of polymer size),
$\beta\delta\mu_c|_{\varrho_c=0} \propto \varrho_p \sigma_c\sigma_p^2$.
For identical physical reasons as in \gl{e5}, the chemical potential 
for inserting a large polymer 
into a sphere fluid becomes intrinsic, $\beta\delta\mu_p|_{\varrho_p=0}\propto
R_g^{1/\nu}/N\propto N^0$ with $\nu=\frac 12$, as the small spheres need to
accommodate 
${\cal O}(R_g^{1/\nu})$  polymer segments. This asymptotic behavior is apparent
in fig. \ref{fig01} and qualitatively agrees with RISM--based 
theory for a single delocalized electron in a hard sphere fluid \cite{leung}. 
 EPS  models, which predict 
$\beta\delta\mu_p|_{\varrho_p=0}\propto
R_g^3/N\propto N^{1/2}$  \cite{Lekkerkerker92}, already begin to deviate from
PRISM for $\xi_c<\sigma_c$, as seen in Fig \ref{fig01}.   
This $N$--dependent overestimate of the polymer insertion 
chemical potential has direct consequences on the predicted trends for 
phase separation as a function of $\xi_c/\sigma_c$,  which are opposite to our
results in fig. \ref{fig02}.  

\begin{figure}[h]
\twoimages[scale=0.38]{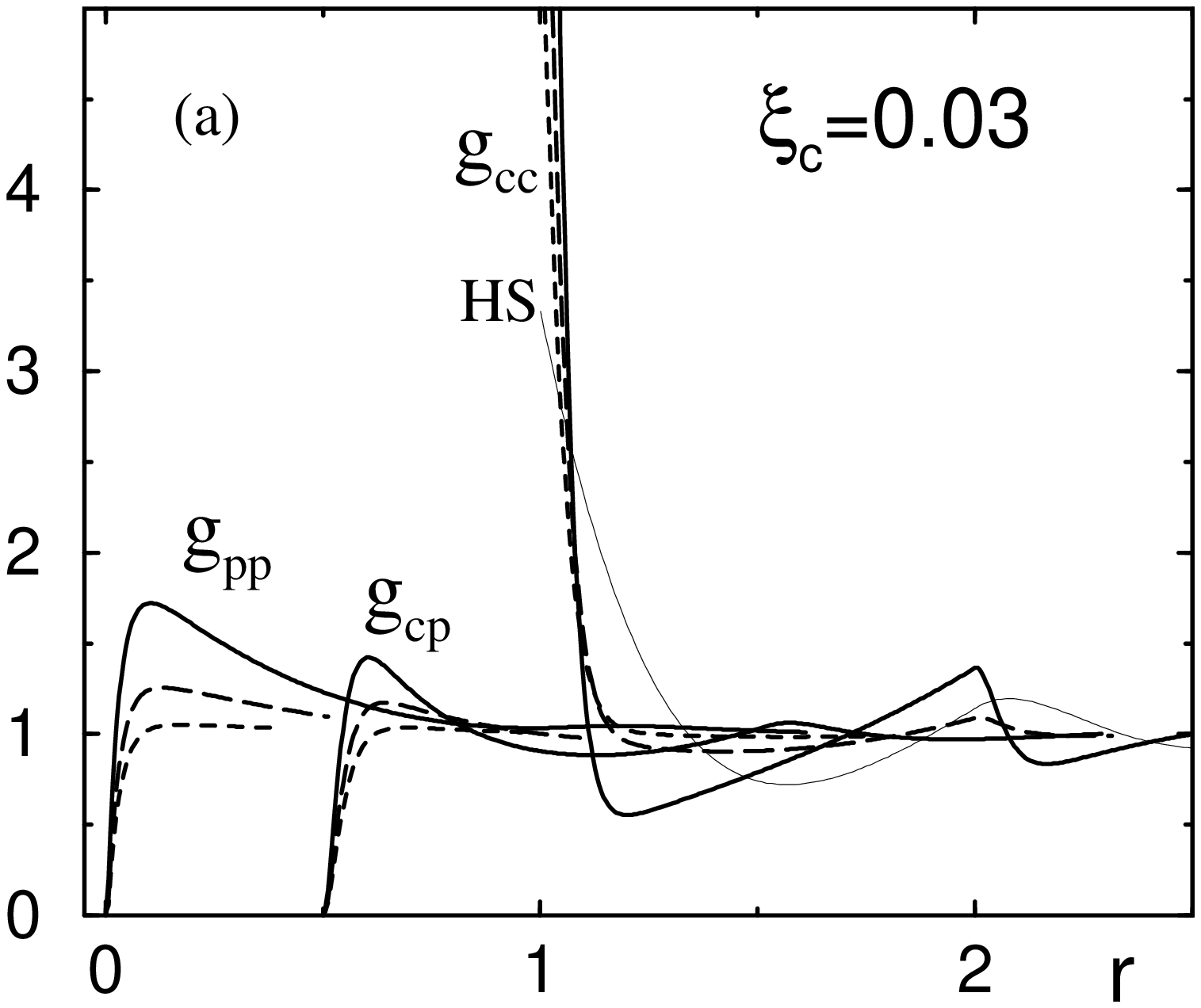}{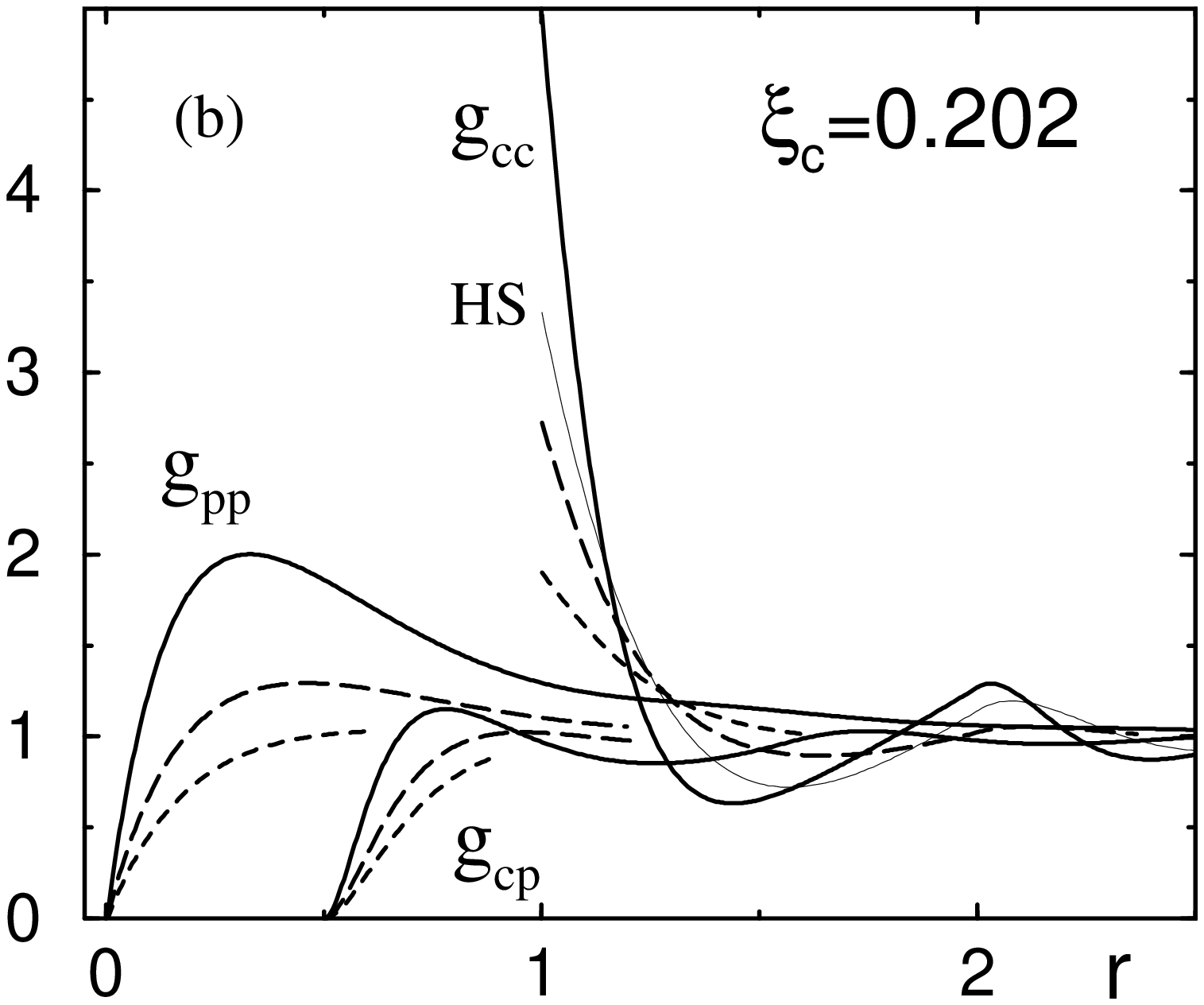}
\caption{Pair correlation functions for two  polymer sizes $\xi_c$.
In (a),  $\xi_c=0.03$ and reduced polymer concentration $\varphi_p=0.061$, 
for increasing colloid packing fractions,
$\phi_c=0.05$, 0.2 and 0.4;
$g_{cc}(r)$ increases to the contact values of $5.80, 8.11$ and
$14.1$.
 In (b),  $\xi_c=0.202$ and identical $\phi_c$ are shown 
at $\varphi_p=0.086$. For both polymer sizes
the small--$q$ limit of the colloid structure factor agrees at $\phi_c=0.40$,
where $S_{cc}(q=0)/ \varrho_c=0.80$. The thin
solid lines in both panels 
labeled HS give the PY--result for hard spheres at $\phi_c=0.40$.
\label{fig1}}
\end{figure}

Figure \ref{fig1} shows the pair correlation functions for different parameter
sets with increasing colloid  
concentrations.
For rather short polymers, fig. \ref{fig1}(a), the polymer induced depletion
attraction leads to  
a strong increase of the colloid contact value, $g_{cc}(r=\sigma_c+)$, relative
to the hard sphere value. 
 The polymer depletion layer is visible in $g_{cp}(r)$ for $r-\frac12<\xi$.
Outside of the depletion region oscillatory correlations are found at higher
$\phi_c$ reflecting the imprinting of colloid order on segment--particle
packing.  Clustering results in the interchain polymer segment
correlations, $g_{pp}(r)$, and fills the correlation hole, which in pure 
dense polymer fluids  results from effective intermolecular
repulsions extending out to $\xi_c$
\cite{kcur3}. 
As the polymer correlations may be considered to decay
for $r\le \sigma_c$, the description of the colloid structure when
$\sigma_c\gg\xi_c$ appears  amenable to traditional one--component
effective pair potential approaches \cite{Asakura,Gast98}, which rely on
 the assumption  of negligible long--ranged polymer correlations.

For larger polymers the depletion layer in fig. \ref{fig1}(b) widens  as
expected. Moreover, increasing  colloid volume fraction leads to much
longer--ranged polymer fluctuations as strikingly apparent 
for $g_{pp}$ at $\phi_c=0.40$. The clustering enforced by the void structure
among the colloidal spheres  overwhelms the polymer repulsion  responsible for
the correlation hole. 
As the polymer correlations extend beyond a number of
particle diameters, effective pair potentials cannot be used.
 Three-- and higher  body contributions to the effective colloid
interactions become important which are mediated by the increasingly correlated
polymer structure; this is in qualitative agreement with simulation studies of
a simplified model using interpenetrating polymers \cite{frenkelprl},  
which should apply to our systems for low polymer concentrations (fugacities).
The  small (large) change of the local structure in $g_{cc}$ for large (small)
polymers,  respectively, is also seen in these simulations.
Our results at higher polymer concentrations further show that the polymer
induced depletion attraction is decreased    
by polymer--polymer excluded volume interactions as has been observed in
neutron scattering 
experiments \cite{tong}. 

\begin{figure}
\onefigure[scale=0.43]{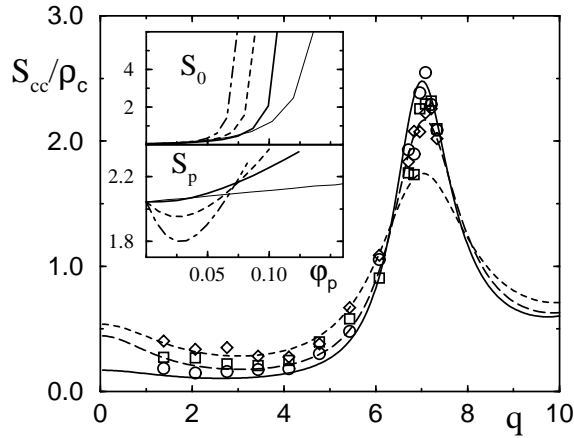}
\caption{
Colloid structure factors compared to  the experimental data by Moussaid
et al. \protect\cite{Moussaid99}. The parameters are
$(\phi_c,\varphi_p,\xi_c)=$ (0.333, 0.069, 0.085; short dashes and $\diamond$),
(0.404, 0.069, 0.131; long dashes and ${\scriptstyle []}$), and
(0.444, 0.053, 0.202; solid line and $\circ$). The inset shows the reduced
polymer concentration dependence 
of the normalized colloid structure factors at the peak
$S_p=S_{cc}(q_p)/\varrho_c$, and at small wave vectors,
$S_0=S_{cc}(0)/\varrho_c$, at fixed colloid packing fraction,
$\phi_c=0.40$, and for increasing polymer sizes,  $\xi_c=$ 0.03 (dot--dashed),
0.085 (short dashed), 0.202 (solid), and 1 (thin solid line).
\label{fig3}}
\end{figure}

The structure of the colloidal liquid at triple coexistence has been the focus
of a recent 
light scattering study which determined $S_{cc}(q)$ for $1.4\le
q\sigma_c\le7.2$ \cite{Moussaid99}. 
Figure \ref{fig3}
 shows the data and theoretical curves evaluated without adjustable
parameter.   The small--$q$ colloid structure shows a strong
dependence on  polymer size. The theory accurately predicts the long wave
length fluctuations which correspond to an order of magnitude enhancement of
the osmotic compressibility relative to pure hard spheres.
The large angle colloid structure experimentally shows little discernible
polymer size dependence. The theory
correctly captures the large--$q$ trend  concerning the peak intensity and 
location, except for the smallest polymer where a larger decrease in the peak
height is predicted  than seen in the 
experiments.  The insets in fig. \ref{fig3} contrast the different  trends of
the colloid structure at small and large wave vectors when changing the
polymer concentration for different polymer sizes. Whereas, the
colloidal osmotic compressibility, $S_{cc}(q=0)/\varrho_c$, increases mainly
due to approaching the spinodal, non--monotonic variations in the large angle
scattering peak height, $S_{cc}(q_p)$, arise for small polymers corresponding
to  a polymer mediated distortion of the local collective packing (``cage'')
around a particle. Based on dynamic mode coupling ideas \cite{mct}
we expect this effect to cause the melting of the
colloidal glass upon addition of small amounts of small polymer ($\xi_c=0.03$) 
as reported in \cite{Ilett95}.

In the limit where the  spheres act as small depletants for the 
larger polymers, the description of the intermolecular structure functions
simplifies. Except for corrections of ${\cal O}(\sigma_c/\xi_c)$, the packing
of the colloids becomes hard sphere like. 
The polymer segment--segment pair correlation function becomes long ranged,
\begin{eqnarray}\label{large}
& g_{pp}(r)\to 1+\frac{\xi_c}{2\varphi_p r}( e^{-\frac{r}{\xi_*}} -
e^{-\frac{r}{\xi_c}})\, \qquad\mbox{for }\; \xi_c\gg\sigma_c\; ,&
\end{eqnarray}
where the collective correlation length, $\xi_*$ with
$\xi_c/\xi_*=1+2\varphi_p(1-f_\infty(\phi_c))$, describes the decay of the
polymer density from its (intermolecular) contact value, $f_\infty(\phi_c)=
\frac{\phi_c(6+\lambda_1-4\lambda_1\phi_c)}{2\lambda_1(1-\phi_c)(1+2\phi_c)}$,
to zero. 
Polymer phase separation can be brought about
by increasing the colloid density, as a spinodal instability 
($1/\xi_*=0$) exists. 
This phase separation  requires finite polymer concentrations,
leads to a dense phase of interpenetrating polymers,  and results from the
increased intermolecular attraction induced by the spherical depletants. 

To summarize, building upon the successful description of the structure of
hard--spheres and  athermal  polymer solutions, we have developed a microscopic
theory of the thermodynamics and structural 
correlations of  binary particle--polymer mixtures. The inclusion of many--body
interaction effects proves crucial for a successful description of the
CPM--structures at 
higher densities. 
The explicit consideration of the conformational entropic contributions of the 
polymers is required to address the packing of larger polymers into the
void space between colloidal spheres. Whereas solely entropic (steric) effects
are considered extensions to
include temperature dependent solvent quality effects and attractions have been
achieved 
for dilute systems \cite{Kulkarni}, and can in principle also be incorporated
for finite concentrations.

Valuable discussions with E. Bartsch,  A. Chatterjee, S. Egelhaaf,
E. Eisenriegler, A. Gast, A. Moussa{\"i}d, W. Poon, P. Pusey and  C. Zukoski
are gratefully acknowledged. M.~F. was supported by the Deutsche
Forschungsgemeinschaft under Grant No. Fu 309/3.
K.~S.~S. was supported by the U.S. DOE Division of Materials Science Grant
No. DEFG02-96ER45539. 


\begin{thebibliography}{10}
\bibitem{Asakura}
\Name{Asakura A. and Oosawa F.}
\REVIEW{J. Chem. Phys.}{22}{1954}{1255}; 
\REVIEW{J. Polym. Sci.}{33}{1958}{183}

\bibitem{Gast98}
\Name{Gast A. P., Hall C. K. and Russel W. B.}
\REVIEW{J. Colloid Interface Sci.}{96}{1983}{251} 

\bibitem{Ilett95}
\Name{Ilett S. M., Orock A., Poon W. C. K. and  Pusey P. N.}
\REVIEW{Phys. Rev. E}{51}{1995}{1344}

\bibitem{zukoski}
\Name{Rosenbaum D., Zamora P. C. and Zukoski C. F.}
\REVIEW{Phys. Rev. Lett.}{76}{1996}{150} 

\bibitem{Poon97}
\Name{Poon W. C. K.}
\REVIEW{Phys. Rev. E}{55}{1997}{3762} 

\bibitem{eisenriegler}
\Name{Eisenriegler E.,  Hanke A. and  Dietrich S.}
\REVIEW{Phys. Rev. E}{54}{1996}{1134} 

\bibitem{joanny}
\Name{Joanny J.-F., Leibler L. and de~Gennes P. G.}
\REVIEW{J. Polymer Sci.: Polymer Physics}{17}{1979}{1073} 

\bibitem{Lekkerkerker92}
\Name{Lekkerkerker H. N. W., Poon W. C. K., Pusey P. N., Stroobants A. and
Warren P. B.} 
\REVIEW{ Europhys. Lett}{20}{1992}{559} 

\bibitem{frenkelprl}
\Name{Meijer E. J. and Frenkel D.}
\REVIEW{J. Chem. Phys.}{100}{1994}{6873};
\REVIEW{Phys. Rev. Lett.}{67}{1991}{1110};
\REVIEW{Physica A}{213}{1995}{130}

\bibitem{tong}
\Name{Ye X., Narayanan T., Tong P. and Huang J. S.}
\REVIEW{Phys. Rev. Lett.}{76}{1996}{4640} 

\bibitem{Moussaid99}
\Name{Moussa{\"i}d A., Poon W.C.K., Pusey P.N. and Soliva M.F.}
\REVIEW{Phys.Rev.Lett.}{82}{1999}{225} 

\bibitem{Louis99}
\Name{Louis A. A.,    Finken R. and Hansen J.-P.}
\REVIEW{Europhys. Lett.}{46}{1999}{741};
\Name{Dijkstra M., Brader J. M. and Evans R.}
\REVIEW{J. Phys.: Condens. Matter}{11}{1999}{10079} 

\bibitem{kcur3}
\Name{Schweizer K. S.  and Curro J. G.}
\REVIEW{ Adv. Chem. Phys}{98}{1997}{1} 

\bibitem{chandl}
\Name{Chandler D.}
\Book{Studies in Statistical Mechanics}
\Editor{Montroll  E.~W. and Lebowitz J. L.}
\Publ{North--Holland, Amsterdam} 
\Year{1982}
\Vol{VIII}
\Page{274}

\bibitem{Chandler93}
\Name{Chandler D.}
\REVIEW{Phys. Rev. E}{48}{1993}{2898} 

\bibitem{threada}
\Name{Fuchs M.}
\REVIEW{Z. Phys. B}{103}{1997}{521} 

\bibitem{footnote} In the limit, $\phi_c\to0$, $\sigma_c\gg\xi_c$ 
 the depletion attraction becomes much larger than $k_BT$ even if
$\varphi_p\ll1$,  \cite{Asakura,joanny}. In this  regime,  
many particle approaches, both classic
\cite{Gast98,Lekkerkerker92} and liquid state theory \cite{avik98},
underestimate the attraction. Our spinodal curves become less reliable
there. 

\bibitem{spinodal} See for example:
\Name{Verhaegh N. A. M., van Duijneveldt J. S.,  Dhont J. K. G. and
Lekkerkerker H. N. W.}
\REVIEW{Physica A}{230}{1996}{409};
\Name{Schaink H. M. and Smit J. A. M.}
\REVIEW{J. Chem. Phys.}{107}{1997}{1004};
\Name{Sperry P. R.}
\REVIEW{J. Colloid Interface Sci.}{99}{1984}{97};
\Name{Robb I. D., Williams P. A., Warren P. and Tanaka R.}
\REVIEW{J. Chem. Soc. Faraday Trans.}{91}{1995}{3901}

\bibitem{avik98}
\Name{Chatterjee A. P. and Schweizer K. S.}
\REVIEW{ J. Chem. Phys.}{109}{1998}{10477} 

\bibitem{leung} 
\Name{Leung K. and Chandler D.}
\REVIEW{ Phys. Rev. E}{49}{1994}{2851} 

\bibitem{mct}
\Name{Fabbian L., G{\"o}tze W., Sciortino F. and Thiery P. T. F}
\REVIEW{Phys. Rev. E}{59}{1999}{R1347};
\Name{Bergenholtz J. and Fuchs M.}
\REVIEW{Phys. Rev. E}{59}{1999}{5706}


\bibitem{Kulkarni}
\Name{Kulkarni A. M., Chatterjee A. P., Schweizer K. S. and Zukoski C. F.}
\REVIEW{Phys.  Rev. Lett.}{83}{1999}{4554};
\Name{Chatterjee A. P. and Schweizer K. S.}
\REVIEW{Macromolecules}{32}{1999}{923}
\end{thebibliography}
\end{document}